\documentclass[showpacs,preprintnumbers,amsmath,amssymb,aps,prl,twocolumn]{revtex4-1}

\usepackage{amsmath,amssymb}
\usepackage[dvips,colorlinks=true,bookmarks=false,citecolor=blue,urlcolor=blue]{hyperref} %latex w/dvips
\usepackage{graphicx}% Include figure files
\usepackage{dcolumn}% Align table columns on decimal point
\usepackage{bm}% bold math
\usepackage{color}
\usepackage{times}
\usepackage{wasysym}
\usepackage{times}
\usepackage{multirow}
\usepackage{verbatim}

\newcommand{\vecr}{\ensuremath{\mathbf{r}}}

\newcommand{\w}{\ensuremath{\omega}}
\newcommand{\e}{\ensuremath{\epsilon}}
\newcommand{\epsbar}{\ensuremath{\bar{\epsilon}}}
\newcommand{\lp}{\ensuremath{\left(}}
\newcommand{\rp}{\ensuremath{\right)}}

\renewcommand{\vec}[1]{\mathbf{#1}}

\renewcommand{\eqref}[1]{(\ref{eq:#1})}
\newcommand{\eqreftwo}[2]{(\ref{eq:#1}), (\ref{eq:#2})}
\newcommand{\Eqref}[1]{Equation~\ref{eq:#1}}
\newcommand{\figref}[1]{Fig.~\ref{fig:#1}}
\newcommand{\Figref}[1]{Figure~\ref{fig:#1}}

\newcommand{\citeasnoun}[1]{Ref.~\onlinecite{#1}}

\newcommand{\eqssref}[2]{Eqs.~(\ref{eq:#1}-\ref{eq:#2})}

\begin{document}

\title{Inverse design of third-order Dirac exceptional points in
  photonic crystals}

\author{Zin Lin$^1$}
\email{zinlin@g.harvard.edu}
\author{Adi Pick$^{2,3}$}
\author{Marko Lon\v{c}ar$^1$}
\author{Alejandro W. Rodriguez$^{4}$}
\affiliation{$^1$John A. Paulson School of Engineering and Applied Sciences, Harvard University, Cambridge, MA 02138}
\affiliation{$^2$Department of Physics, Harvard University, Cambridge, MA 02138}
\affiliation{$^3$Department of Mathematics, Massachusetts Institute of Technology, Cambridge, MA 02139}
\affiliation{$^4$Department of Electrical Engineering, Princeton University, Princeton, NJ, 08544}

\date{\today}

\begin{abstract}
  We propose a novel inverse-design method that enables brute-force
  discovery of photonic crystal (PhC) structures with complex spectral
  degeneracies. As a proof of principle, we demonstrate PhCs
  exhibiting third-order Dirac points formed by the \emph{accidental}
  degeneracy of modes of monopolar, dipolar, and quadrupolar
  nature. We show that under suitable conditions, these modes can
  coalesce and form a third-order exceptional point (EP3), leading to
  diverging Petermann factors. We show that the spontaneous emission
  (SE) rate of emitters at such EP3s, related to the local density of
  states, can be enhanced by a factor of 8 in purely lossy (passive)
  structures, with larger enhancements $\sim \sqrt{n^3}$ possible at
  exceptional points of higher order $n$ or in materials with gain.
\end{abstract}

\pacs{Valid PACS appear here}%
\maketitle

Dirac cones in photonic systems have received much attention because
of their connections to intriguing optical properties, enabling
large-area photonic-crystal (PhC) surface-emitting
lasers~\cite{MarinDirac}, zitterbewegung of photons~\cite{Zhang08},
appearance of zero-index behavior~\cite{CTChan11, Yang15}, and as
precursors to nontrivial topological effects~\cite{Haldane, MacDonald,
  LingLu}. Recent work also showed that Dirac-point degeneracies can
give rise to rings of exceptional points~\cite{Zhen15}. An exceptional
point (EP) is a singularity in a non-Hermitian system where two or
more eigenvectors and their corresponding complex eigenvalues
coalesce, leading to a non-diagonalizable, defective
Hamiltonian~\cite{Moiseyev11, Kato95}. EPs have been studied in
various physical contexts, most notably lasers and atomic as well as
molecular systems~\cite{Berry04, Heiss12}. In recent decades, interest
in EPs has been re-ignited in connection with non-Hermitian
parity-time symmetric systems~\cite{Bender98}, especially optical
media involving carefully designed gain and loss
profiles~\cite{Ruter10, Guo09, Mei10, Hamid12, Longhi14, Ge14,
  Cerjan16}, where they can lead to intriguing phenomena such as
enhanced spontaneous emission (SE)~\cite{Berry03, Adi16}, chiral
modes~\cite{Dembowski03}, directional transport~\cite{Zin11, Feng13}
and anomalous lasing behavior~\cite{Liertzer12, Hodaei14,
  Feng14}. Also recently, it became possible to directly observe EPs
in photonic crystals (PhC)~\cite{Zhen15} and optoelectronic
microcavities~\cite{Gao15}. Thus far, however, only second-order EPs
(EP2) (where only two modes coalesce) have been proposed in the
context of photonic radiators: in fact, apart from a few mathematical
analyses~\cite{Graefe08, Ryu12, Wunner15} or very recently, acoustic
systems~\cite{PRX16}, there has been little or no investigation into
appearance of EPs of higher order (where more than two modes collapse)
in complex photonic geometries.

In this letter, we propose a powerful inverse-design method based on
topology optimization that allows automatic discovery of complex
photonic structures supporting Dirac points (DP) formed out of the
\emph{accidental} degeneracy~\footnote{By accidental degeneracy, we
  mean that the frequency collision is neither induced nor protected
  by any underlying point symmetry, but rather is entirely fashioned
  out of the detailed morphology of the photonic unit cell, designed
  via brute-force topology optimization techniques.} of modes
belonging to different symmetry representations. In particular, we
show that such higher-order DPs can be exploited to create third-order
exceptional points (EP3) along with complex contours of EP2.  In
addition, we exploit coupled-mode theory to derive conditions under
which such EP3s can exist and extend recent work~\cite{Adi16} to
consider the possible enhancements and spectral modifications in the
SE rate of emitters. Specifically, we show that the local density of
states at a EP3 can be enhanced 8-fold (in passive systems) and can
exhibit a cubic Lorentzian spectrum under special conditions.  More
generally, we find that the enhancement factor $\sim \sqrt{n^3}$ with
increasing EP order $n$, whilst even larger enhancements are expected
under gain~\cite{Adi16}. Our findings provide the foundations for
future discoveries of complex structures with unusual or exotic modal
properties currently out of the reach of conventional, intuitive
design principles.

Dirac cones and Dirac EPs are typically designed by exploiting
degeneracies between modes of different symmetry representations,
often in simple geometries involving cylindrical pillars or holes on a
square or triangular lattice~\cite{CTChan11, CTChan12}. These
singularities are typically of order two (comprising two interacting
modes) and arise partly out of some underlying lattice symmetry
(e.g. $C_{4v}$ or $C_{3v}$) and through the fine-tuning of a few
geometric parameters~\cite{CTChan11, Sakoda12}. For instance,
in~\citeasnoun{Zhen15}, it was recently demonstrated that a Dirac
point (DP) at the $\Gamma$ point of a PhC with $C_{4v}$ symmetry can
give rise to a ring of EP2s. Such a DP is formed by a degeneracy
involving modes of both monopolar (M) and dipolar (D) nature, which
transform according to $A$ and $E$ representations of the $C_{4v}$
group~\cite{CTChan11, Sakoda12}. Even though the degeneracy consists
of one monopole and two dipoles, the induced EP is of the second
order, with only the monopole and one of the dipoles colliding, while
the coalescence of the dipole partner is prevented by their
symmetry~\cite{Zhen15}. Below, we show that an EP3 can be induced by a
completely ``accidental'' third-order degeneracy (D3) at $\Gamma$,
involving modes of monopolar (M), dipolar (D) and quadrupolar (Q)
nature arising in a novel, inverse-designed PhC structure lacking
$C_{4v}$ symmetry.

{\it Coupled-mode analysis.---} The band structure in the vicinity of
such a D3 can be modeled by an approximate Hamiltonian of the
form~\cite{CTChan12}:
\begin{align}
\mathcal{H}=
\begin{pmatrix}
\omega_0 & v_\text{MD} k_x & 0 \\
v_\text{MD} k_x & \omega_0 & v_\text{QD} k_y \\
0 & v_\text{QD} k_y & \omega_0 
\end{pmatrix}
\end{align}
Here, $v_{ij},~i,j\in\{\mathrm{M, D, Q}\}$ characterizes the mode
mixing away from the $\Gamma$ point, to first order in
$\vec{k}$~\cite{CTChan12}. Note that the diagonalization of this
Hamiltonian yields a completely real band structure comprising a Dirac
cone and a flat band,
\begin{align}
\omega=\omega_0,~\omega_0 \pm 
\sqrt{v_\text{MD}^2 k_x^2 + v_\text{QD}^2 k_y^2}
\end{align}

To induce an EP, non-Hermiticity can be introduced by the addition of
a small imaginary perturbation to the Hamiltonian,
\begin{align}
\mathcal{H}=
\begin{pmatrix}
\omega_0 + i\gamma_\text{M} & v_\text{MD} k_x & 0 \\
v_\text{MD} k_x & \omega_0 + i\gamma_\text{D} & v_\text{QD} k_y \\
0 & v_\text{QD} k_y & \omega_0 + i \gamma_\text{Q} 
\end{pmatrix}
\label{eq:H}
\end{align}
with $\gamma > 0 $ $(< 0)$ representing a small amount of absorption
(amplification) or radiation. A EP3 is obtained by demanding that the
characteristic polynomial of \eqref{H} have vanishing derivatives up
to second order,
\begin{align}
\label{eq:det1}
  \det{\left( \mathcal{H} - \omega \mathbb{I} \right)} = P(\omega) &= 0, \\
  P'(\omega) &=0, \\
  P''(\omega) &= 0.
\label{eq:det2}
\end{align}
Solving the above equations for $\omega,~k_x$, and $k_y$ yields the
EP3:
\begin{align}
  \omega^\text{EP3} &= \omega_0 + {i \over 3} \left( \gamma_\text{M} + \gamma_\text{D} + \gamma_\text{Q} \right) \\
  k_x^\text{EP3} &= \pm {1 \over 3 v_\text{MD}} \sqrt{ \left( \gamma_\text{D} + \gamma_\text{Q} - 2 \gamma_\text{M} \right)^3 \over 3 \left( \gamma_\text{Q} - \gamma_\text{M} \right)  } \label{eq:kepx}\\
  k_y^\text{EP3} &= \pm {1 \over 3 v_\text{QD}} \sqrt{ \left(
      2\gamma_\text{Q} - \gamma_\text{M} - \gamma_\text{D} \right)^3
    \over 3 \left( \gamma_\text{Q} - \gamma_\text{M} \right) } \label{eq:kepy}
\end{align}
where, any choice of distinct $\gamma$ leading to real $\vec{k}$
induces an EP3.  In a lattice with $C_{4v}$ symmetry, this condition
cannot be satisfied, unless the symmetry relating the two dipolar
modes is severely and intentionally broken. Such a design would
necessitate an overlay of spatially varying regions of gain/loss, a
scenario that seems experimentally challenging. In contrast, we now
present a novel design method that can discover PhC geometries
supporting ``accidental'' and tunable D3s.

\begin{figure}[t!]
\centering
\includegraphics[width=1\columnwidth]{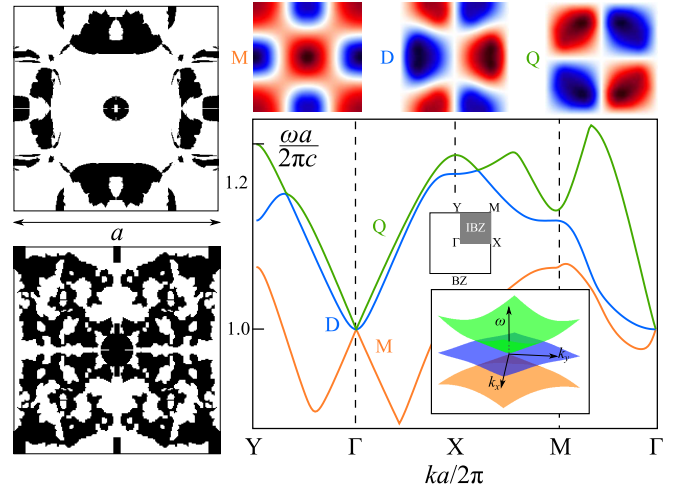}
\caption{Inverse-designed 2d square lattices comprising either
  low-index ($n=2$, upper left) or high-index ($n=3$, lower left)
  materials in air (white regions), with periodicity $a/\lambda=1.05$
  and $a=0.6/\lambda$, respectively. Note that both unit cells lack
  $C_{4v}$ symmetry, but retain $C_{2v}$ symmetry by design. Lower
  right: band structure of the low-index (upper) lattice, revealing a
  Dirac point induced by the presence of an \emph{accidental}
  third-order degeneracy (D3) of monopolar (M), dipolar (D), and
  quadrupolar (Q) modes (upper insets), leading to linear Dirac
  dispersion accompanied by a quadratic flat band at the $\Gamma$
  point. A schematic of the Brillouin zone (BZ) denoting high-symmetry
  $\mathbf{k}$ points $(\mathrm{Y,\Gamma,X,M})$ is also shown. Due to
  the lack of $C_{4v}$ symmetry, the dispersion along the X and Y
  directions differ.~\label{fig:fig1}}
\end{figure}

{\it Inverse-design method.---} We construct an accidental D3 by
employing a large-scale optimization strategy for automatically
discovering novel topologies and geometries impossible to conceive
from conventional intuition alone. One such strategy, known as
topology optimization (TO), employs powerful gradient-based numerical
algorithms capable of handling a very large design space, typically
considering every pixel or voxel as a degree of freedom (DOF) in an
extensive 2d or 3d computational domain. Such techniques have been
gaining traction and were recently applied to problems involving
linear input/output coupling of
light~\cite{Jensen11,Piggott15,Shen15}, cavity Purcell
enhancement~\cite{Liang13}, and nonlinear frequency
conversion~\cite{Zin16}. In this work, we apply TO to the problem of
inverse-designing the band structure of a PhC to support spectral DP
degeneracies and EP singularities.

\begin{figure*}[t]
\centering
\includegraphics[width=1.5\columnwidth]{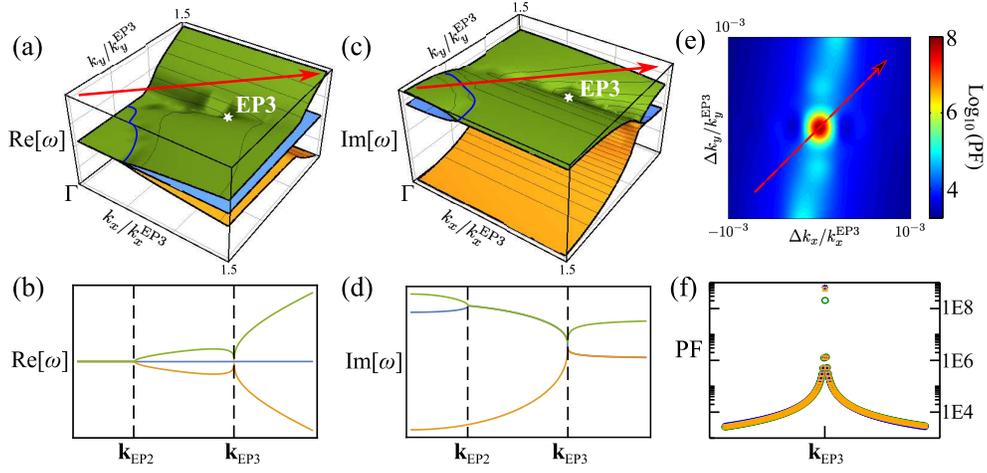}
\caption{(a) Real and (c) imaginary eigenfrequencies as a function of
  $k_x$ and $k_y$ in the vicinity of a third order exceptional point
  (EP3) of the structure described in \figref{fig1}, located at
  $\mathbf{k}^\text{EP3}\approx\{7,1.8\}\times 10^{-5}~(\frac{2
    \pi}{a})$ (red dot). The blue contours denote regions of
  second-order exceptional points where two of the three modes
  coalesce. The plots in (b) and (d) show the corresponding band
  structures along the $\vec{k}$ lines marked by red arrows. (e)
  Contour plot showing the enhanced Petermann factor (PF) associated
  with one of the modes in the vicinity of the EP3, and (e)
  corresponding enhancement along the direction shown by the red arrow,
  for all three modes.~\label{fig:fig2}}
\end{figure*}

Our approach extends the work of \citeasnoun{Liang13}, which showed
that it is possible to design a structure supporting a resonant mode
at some arbitrary frequency by maximizing the time-averaged power
output $f= -\mathrm{Re} \Big[ \int
\mathbf{J}^*\cdot\mathbf{E}~d\mathbf{r} \Big]$ emitted from a time
harmonic current source $\mathbf{J}$ at the desired frequency
$\omega$, where the electric field response $\mathbf{E}$ is given by
the solution of Maxwell's equations, $\nabla \times {1 \over \mu}
\nabla \times \mathbf{E} - \omega^2 \epsilon(\mathbf{r}) \mathbf{E} =
i \omega \mathbf{J}$~\cite{Liang13}. To ensure that the designed
resonance has the requisite modal profile, the current $\mathbf{J}$
must be judiciously constructed. For example, to design a transverse
magnetic (TM) polarized monopolar mode (M) at the $\Gamma$ point of a
PhC, $\mathbf{J}$ should can be chosen as a point dipole $\mathbf{J} =
\delta(\mathbf{r} - \mathbf{r}_0) \mathbf{e}_z$ at the center
$\mathbf{r}_0$ of the unit cell. Once the objective function $f$ is
identified, its gradient with respective to $\epsilon(\mathbf{r})$ can
be calculated by the so-called adjoint variable method~\cite{Jensen11,
  Liang13} (see the supplement for details) and then supplied to any
large-scale gradient-based optimization algorithm such as the method
of moving asymptotes (MMA)~\cite{Svanberg02}. To design structures
supporting multiple modes at the same frequency with the requisite (M,
D, Q) symmetries, we seek a maxmin formulation in which one maximizes
the minimum of $\{f_\text{M},~f_\text{D},~f_\text{Q}\}$, with currents
chosen to ensure fields with the desired symmetries, discussed in
detail in the supplementary materials [SM].

Our topology optimization framework can be exploited to design
high-order degeneracies with distinct modal properties in arbitrary
material systems and photonic structures. Here, we use it to
demonstrate the appearance of third-order degeneracies in binary
dielectric/air square lattices. \Figref{fig1}(left) shows two such
structures, involving materials of either low ($n=2$, upper) or high
($n=3$, lower) refractive indices (in air) and periodicities
$a=1.05~\lambda$ and $a=0.6~\lambda$, respectively, where $\lambda$ is
the design wavelength in vacuum. Note that such refractive indices are
typical for common materials such as silicon nitride, lithium niobate,
diamond, silicon, alumina, or many low and high-index ceramics at
optical, microwave, and terahertz frequencies.  We focus our
discussion on the low-index structure, leaving details of the
high-index design to the [SM]. Noticeably, the band structure of the
low-index lattice exhibits a D3 comprising M, D and Q modes at the
$\Gamma$ point, shown in \figref{fig1} (lower right). Note that since
the optimized PhC lacks $C_{4v}$ symmetry (but possesses $C_{2v}$),
there is only one dipolar mode at the designated frequency and hence,
the degeneracy of the three modes is completely \emph{accidental}:
potential mode mixing and avoided crossings at the $\Gamma$ point are
prevented by the corresponding mirror symmetries. In the vicinity of
the tri-modal degeneracy, the band structure exhibits conical Dirac
dispersion accompanied by a quadratic flat band.  While general rules
regarding the occurrence of Dirac point (DP) dispersion in the
vicinity of a modal degeneracy are well understood from group
theoretic considerations, e.g. as arising from \emph{two} different
irreducible representations~\cite{Sakoda12}, to our knowledge our
TO-designed PhC is the first demonstration of a DP formed by
\emph{three} degenerate modes belonging to \emph{three} different
representations, namely the A$_1$, A$_2$ and B$_1$ representations of
the $C_{2v}$ group.

%or which could be realized by a variety of material processing methods
%such as doping, controlled impurities, metallic depositions and

\begin{figure*}[t]
\centering
\includegraphics[width=2.05\columnwidth]{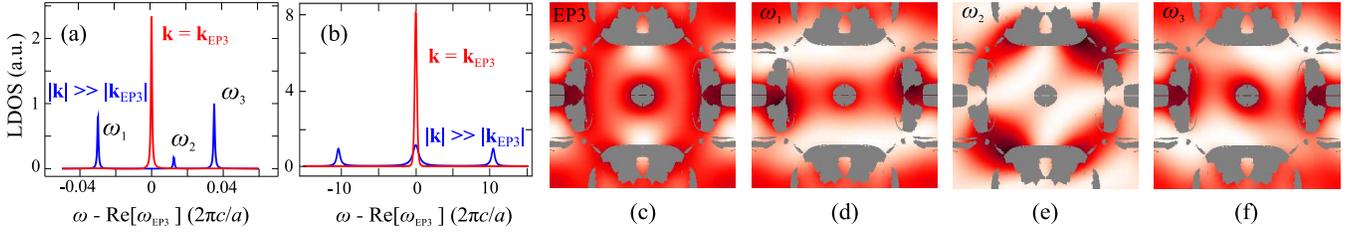}
\caption{(a) Local density of states (LDOS) at the center of the unit
  cell of the structure in \figref{fig1}, evaluated at either
  $\mathbf{k}^\text{EP3} \approx \{7,1.8\}\times 10^{-5}
  (\frac{2\pi}{a})$ (red curves) or $\mathbf{k} = \{7,1.8\}\times
  10^{-2} (\frac{2\pi}{a}) \gg \mathbf{k}^\text{EP3}$ (blue
  curves). (b) Maximum (8-fold) LDOS enhancement associated with a
  EP3, computed via the reduced $3 \times 3$ Hamiltonian model
  of~\eqref{H}. (c)--(f) LDOS profiles evaluated at either
  $\omega_\text{EP3}$ or at the non-degenerate frequencies
  $\omega_1,~\omega_2$, and $\omega_3$, corresponding to the EP3 and
  far-away points described in (a). Note that the LDOS is evaluated
  only in air regions since the LDOS within a lossy medium formally
  diverges~\cite{Scheel99}.~\label{fig:fig3}}
\end{figure*}

{\it Third-order exceptional point.---} The third order Dirac
degeneracy of \figref{fig1} can be straightforwardly linked to an EP3
through the introduction of non-Hermiticity, i.e.  material loss,
gain, or open boundaries (radiation). Here, we consider such an EP3 by
introducing a small imaginary part in the dielectric constant, $\kappa
= \sqrt{\mathrm{Im}[\epsilon]} = 0.005$, representing intrinsic
material loss and resulting in small decay rates
$\{\gamma_\text{M},\gamma_\text{D},\gamma_\text{Q}\}/\omega_0 \approx
\{3.6,4.3,4.2\} \times 10^{-4}$. From \eqreftwo{kepx}{kepy}, it
follows that there exists an EP3 at $\mathrm{Re}[\omega_\text{EP3}]
\approx \omega_0$, $\mathrm{Im}[\omega_\text{EP3}] \approx 4 \times
10^{-4}({2 \pi c \over a})$, $k_x^\text{EP3} \approx 7\times
10^{-5}~({2 \pi \over a})$ and $k_y^\text{EP3} \approx 1.8\times
10^{-5}~({2 \pi \over a})$ [SM]. \Figref{fig2}(a) and (c) show the
band structure in the vicinity of the $\Gamma$ point, along with
slices, \figref{fig2}(b) and (d), indicated by blue arrows,
illustrating the coalescence of both the real and imaginary mode
frequencies. Yet another interesting feature of the dispersion
landscape is that, apart from the EP3, there also exists a contour of
EP2 (blue lines), defined by $P(\omega;k_x,k_y) =
0,~P'(\omega;k_x,k_y)=0$, similar to the ring of EP2 observed
in~\citeasnoun{Zhen15}.

A defining signature of non-Hermitian systems is that eigenvectors are
no longer orthogonal. Rather, they are bi-orthogonal~\cite{Kato95} in
the sense of an unconjugated ``inner product'' between left and right
eigenvectors, $\left(\Psi^\text{L}_n\right)^\text{T} \Psi^\text{R}_m =
\delta_{nm}$, defined such that $A \Psi^\text{R} = \omega^2
\Psi^\text{R}$ and $A^\text{T}\Psi^\text{L} = \omega^2 \Psi^\text{L}$,
where $A$ is the Maxwell operator $\hat{\epsilon}^{-1} (\nabla
+i\vec{k}) \times {1\over \mu} (\nabla+i\vec{k}) \times$ under Bloch
boundary conditions at a specific $\mathbf{k}$, $\hat{\epsilon}$ is
the diagonal permittivity tensor $\epsilon(\mathbf{r})$. At our EP3 ,
the three eigenmodes coalesce and become self-orthogonal~\cite{Mei10},
leading to vanishing inner products
$\left(\Psi^\text{L}_n\right)^\text{T} \Psi^\text{R}_n = 0,~n\in \{
\mathrm{1,2,3} \mathrm \}$, as characterized by the so-called
Petermann factor (PF),
\begin{align} 
  \mathrm{PF}_n = { ||\Psi^\text{L}_n||^2 ||\Psi^\text{R}_n||^2 
    \over |\left(\Psi^\text{L}_n\right)^\text{T} \Psi^\text{R}_m|^2 }
\end{align}
where $||...||^2$ is the usual L$_2$ norm given by $||\Psi||^2 =
\Psi^{\text{T}*} \Psi$. \Figref{fig2}(e,f) illustrates the divergence
of the PF for all three modes as $\vec{k} \to
\vec{k}^\text{EP3}$. Note that there are also PF divergences
associated with the M, D modes at the EP2 contours.

{\it Local density of states.---} The divergence of the Petermann
Factor (PF) in open systems can lead to many important
effects~\cite{Petermann1979, Berry04}. In particular, the SE rate of
emitters in resonant cavities is traditionally expressed via the PF (a
generalization of the Purcell factor~\cite{Petermann1979}), becoming
most pronounced near EPs where the latter
diverges~\cite{berry2003mode}. More rigorously, however, the SE rate
is given by the local density of states (LDOS), or electromagnetic
Green's function (GF), which though enhanced turns out to be finite
even at EPs~\cite{Adi16}: coalescent eigenmodes no longer form a
complete basis, requiring instead an augmented basis of associated
Jordan modes and hence a different definition of LDOS. Such an
expansion was recently employed in~\citeasnoun{Adi16} to demonstrate
limits to LDOS at EP2s in both passive and active media; here, we
extend these results to the case of EP3s.

The LDOS at an EP3 can be obtained from the diagonal elements of the
imaginary part of the dyadic GF [SM]: 
%\small
\begin{multline}
  \mathbb{G}_\mathrm{EP3} \approx \frac{\Psi_\text{EP3}^\text{R}
    (\Phi_\text{I}^\text{L})^T}{(\omega^2-\omega_\text{EP3}^2)^3}+
  \frac{\Psi_\text{EP3}^\text{R} (\Phi_\text{I}^\text{L})^T +
    \Phi_\text{I}^\text{R}
    (\Psi_\text{EP3}^\text{L})^T}{(\omega^2-\omega_\text{EP3}^2)^2} 
  \\ + \frac{\Psi_\text{EP3}^\text{R} (\Phi_\text{II}^\text{L})^T +
    \Phi_\text{I}^\text{R} (\Phi_\text{I}^\text{L})^T +
    \Phi_\text{II}^\text{R}
    (\Psi_\text{EP3}^\text{L})^T}{\omega^2-\omega_\text{EP3}^2}.
\label{eq:G_ep3}
\end{multline}
\normalsize \Eqref{G_ep3} involves a complicated sum of cubic,
quadratic, and linear Lorentzian profiles weighted by the outer
products of the only surviving left (right) eigenmode
$\Psi_\text{EP3}^\text{(L,R)}$ and the two associated Jordan vectors
$\Phi_\text{(I,II)}^\text{(L,R)}$, determined by the third-order
Jordan decomposition of the Maxwell eigenproblem,
\begin{align}
A_\text{EP3} \Psi_\text{EP3}^\text{R} &= \omega_\text{EP3}^2 \Psi_\text{EP3}^\text{R} \\
A_\text{EP3} \Phi_\text{I}^\text{R} &= \omega_\text{EP3}^2  \Phi_\text{I}^\text{R} + \Psi_\text{EP3}^\text{R} \\
A_\text{EP3} \Phi_\text{II}^\text{R} &= \omega_\text{EP3}^2 \Phi_\text{II}^\text{R} + \Phi_\text{I}^\text{R},
\end{align}
and its associated dual. \Eqref{G_ep3} reveals that the LDOS spectrum
$\sim -\mathrm{Im}\big[\mathrm{Tr} \big(\mathbb{G}\big)\big]$ can vary
dramatically depending on position, frequency, and decay rates.

\Figref{fig3}(a) shows the LDOS spectra at the center of the unit cell
$\vec{r}_0$, evaluated at either $\vec{k}^\text{EP3}$ (red curves) or
a point $\vec{k} = \{7,1.8\} \times 10^{-2} (2 \pi / a) \gg
\vec{k}^\text{EP3}$ (blue curves) far away from the EP3, demonstrating
an enhancement factor of $\approx 2.33$ in this geometry. Even greater
enhancements are possible under different loss profiles, i.e.,
$\gamma_\text{M},~\gamma_\text{D}$ and $\gamma_\text{Q}$, as
illustrated by the following analysis based on the reduced Hamiltonian
framework above. In particular, the GF at a given location in the unit
cell can be directly related to the diagonal entries of the resolvent
of $\mathcal{H}$, defined as $G \equiv (\mathcal{H}-\omega
\mathbb{I})^{-1}$. For example, the third entry of $G$ yields the LDOS
at points where the intensity of the quadrupole mode dominates.
Consider a scenario in which only the monopole mode has a finite
lifetime, i.e., $\gamma_\text{M}=\gamma$ while $\gamma_\text{D} =
\gamma_\text{Q}=0$. It follows from \eqref{H} and \eqref{G_ep3} that
the LDOS in this case is given by,
%\small
\begin{multline}
  -\mathrm{Im}\{G_\text{EP3}[3,3]\} \approx
  -\frac{2\gamma^2}{27}\frac{\bar{\gamma}^3-3\bar{\gamma}(\mathrm{Re}[\omega_\text{EP3}]-\omega)^2}{(\mathrm{Re}[\omega_\text{EP3}]-\omega)^2+\bar{\gamma}^2]^3}\\
  +\frac{\gamma}{3}\frac{\bar{\gamma}^2-(\mathrm{Re}[\omega_\text{EP3}]-\omega)^2}{[(\mathrm{Re}[\omega_\text{EP3}]-\omega)^2+\bar{\gamma}^2]^2}
  -\frac{\bar{\gamma}}{(\mathrm{Re}[\omega_\text{EP3}]-\omega)^2+\bar{\gamma}^2},
\label{eq:LDOSquad}
\end{multline}
\normalsize where $\bar{\gamma} \equiv \gamma/3$. Moreover, the peak
LDOS at $\omega=\mathrm{Re}[\omega_\text{EP3}]$ is found to be
$8/\gamma$, corresponding to an 8-fold enhancement relative to the
peak LDOS far away from the EP3. Such an enhancement is illustrated in
\figref{fig3}(b), which also reveals the highly non-Lorentzian
spectrum associated with this EP3.

It is possible to exploit a simple sum rule, namely that the
spectrally integrated LDOS is a constant~\cite{barnett1996sum}, to
predict the maximum enhancement possible for an EP of arbitrary order
$n$.  In particular, the integrated LDOS of an order-$n$ Lorentzian of
the form $L_n(\omega)=\frac{\gamma^{2n-1}
  c_n}{[(\omega-\mathrm{Re}[\omega_\text{EPn}])^2+\gamma^2]^n}$ is
$S_n(\omega)=\int\!d\omega\,L_n(\omega)=
\frac{c_n\sqrt{\pi}\Gamma[n-\tfrac{1}{2}]}{\Gamma[n]}$, where $\Gamma$
is the gamma function.  It follows from the sum rule that
$nS_1(\omega)=S_n(\omega)$ and, consequently, that
$c_n/c_1=\frac{\sqrt{\pi}\Gamma[n+1]}{\Gamma[n-\tfrac{1}{2}]} \sim
\sqrt{n^3}$ for large $n\gg 1$.  In the case of an EP3, the maximum
enhancement $c_3/c_1 = 8$, which is realized in the scenario discussed
above.

{\it Concluding remarks.---} The inverse-design approach described
above is a powerful, general-purpose tool for engineering complex and
unusual photonic properties, such as spectral degeneracies, leading to
unconventional structures that arguably could not have been conceived
by intuition alone. Although fabrication of the resulting ``bar-code''
structures may prove challenging at visible wavelengths using
currently available technologies, future experimental realizations are
entirely feasible in the far-infrared to microwave regimes, where
complex features can be straightforwardly fabricated in polymers and
ceramics with the aid of computerized machining, 3D printing, laser
cutting, additive manufacturing, or two-photon
lithography~\cite{Borisov98, Lewis15, Vukusic16}. Furthermore, while
our predictions offer a proof of principle based on a particular PhC
platform, the same inverse-design techniques can be applied to
consider higher-order EPs as well as other topologies, including
localized cavities. Our ongoing work in this regard includes
application of TO to problems related to the design of chiral modes,
photonic Weyl points, topological insulators, and omnidirectional
Dirac-cone, zero-index meta-materials.

\emph{Acknowledgments.---} We would like to thank Steven G. Johnson
for useful discussions. This work was partially supported by the Air
Force Office of Scientific Research under contract FA9550-14-1-0389,
by the National Science Foundation under Grant no. DMR-1454836, and by
the Princeton Center for Complex Materials, a MRSEC supported by NSF
Grant DMR 1420541. Z. Lin is supported by the National Science
Foundation Graduate Research Fellowship Program under Grant
No. DGE1144152.

%\tp{When we were preparing this manuscript, we became aware of the
%  publication of a paper~\cite{PRX16} which discusses the realization
%  of higher order exceptional points in acoustic systems. While the
%  work presented in the aforementioned paper has certain similarities
%  to ours in terms of their emphasis on higher order EPs, our work
%  crucially differs from theirs in that we present a unconventional
%  versatile computational technique to design complex spectral
%  degeneracies and coalescent sigularities via topology optimization
%  of extended photonic crystals while their higher order EP design
%  follows the more conventional route of engineering loss and coupling
%  rates among localized resonators. Furthermore, we have also analyzed
%  the LDOS properties of higher order EPs.}

\bibliographystyle{unsrt}
\bibliography{ep3,bibliographyEP3}

\begin{thebibliography}{10}

\bibitem{MarinDirac}
Song-Liang Chua, Ling Lu, Jorge Bravo-Abad, John~D. Joannopoulos, and Marin
  Solja\v{c}i\'{c}.
\newblock Larger-area single-mode photonic crystal surface-emitting lasers
  enabled by an accidental dirac point.
\newblock {\em Opt. Lett.}, 39(7):2072--2075, Apr 2014.

\bibitem{Zhang08}
Xiangdong Zhang.
\newblock Observing \textit{Zitterbewegung} for photons near the dirac point of
  a two-dimensional photonic crystal.
\newblock {\em Phys. Rev. Lett.}, 100:113903, 2008.

\bibitem{CTChan11}
Xueqin Huang, Yun Lai, Zhi~Hong Hang, Huihuo Zheng, and C.~T. Chan.
\newblock Dirac cones induced by accidental degeneracy in photonic crystals and
  zero-refractive-index materials.
\newblock {\em Nat Mater}, 10(8):582--586, 08 2011.

\bibitem{Yang15}
Yang Li, Shota Kita, Philip Mu{\~n}oz, Orad Reshef, Daryl~I. Vulis, Mei Yin,
  Marko Lon{\v c}ar, and Eric Mazur.
\newblock On-chip zero-index metamaterials.
\newblock {\em Nat Photon}, 9(11):738--742, 11 2015.

\bibitem{Haldane}
S.~Raghu and F.~D.~M. Haldane.
\newblock Analogs of quantum-hall-effect edge states in photonic crystals.
\newblock {\em Phys. Rev. A}, 78:033834, 2008.

\bibitem{MacDonald}
Alexander~B. Khanikaev, S.~Hossein~Mousavi, Wang-Kong Tse, Mehdi Kargarian,
  Allan~H. MacDonald, and Gennady Shvets.
\newblock Photonic topological insulators.
\newblock {\em Nat Mater}, 12(3):233--239, 2013.

\bibitem{LingLu}
Ling Lu, Chen Fang, Liang Fu, Steven~G. Johnson, John~D. Joannopoulos, and
  Marin Soljacic.
\newblock Symmetry-protected topological photonic crystal in three dimensions.
\newblock {\em Nat Phys}, 12(4):337--340, 2016.

\bibitem{Zhen15}
Bo~Zhen, Chia~Wei Hsu, Yuichi Igarashi, Ling Lu, Ido Kaminer, Adi Pick,
  Song-Liang Chua, John~D. Joannopoulos, and Marin Soljacic.
\newblock Spawning rings of exceptional points out of dirac cones.
\newblock {\em Nature}, 525(7569):354--358, 09 2015.

\bibitem{Moiseyev11}
Nimrod Moiseyev.
\newblock {\em Non-Hermitian Quantum Mechanics}.
\newblock Cambridge University Press, 2011.

\bibitem{Kato95}
Tosio Kato.
\newblock {\em Perturbation theory for linear operators}.
\newblock Springer-Verlag Berlin Heidelberg, 1995.

\bibitem{Berry04}
Michael~V. Berry.
\newblock Physics of nonhermitian degeneracies.
\newblock {\em Czechoslovak Journal of Physics}, 54(10), 2004.

\bibitem{Heiss12}
W~D Heiss.
\newblock The physics of exceptional points.
\newblock {\em Journal of Physics A: Mathematical and Theoretical},
  45(44):444016, 2012.

\bibitem{Bender98}
Carl~M. Bender and Stefan Boettcher.
\newblock Real spectra in non-hermitian hamiltonians having
  $\mathcal{P}\mathcal{T}$ symmetry.
\newblock {\em Phys. Rev. Lett.}, 80:5243--5246, 1998.

\bibitem{Ruter10}
Christian~E. Ruter, Konstantinos~G. Makris, Ramy El-Ganainy, Demetrios~N.
  Christodoulides, Mordechai Segev, and Detlef Kip.
\newblock Observation of parity-time symmetry in optics.
\newblock {\em Nat Phys}, 6(3):192--195, 2010.

\bibitem{Guo09}
A.~Guo, G.~J. Salamo, D.~Duchesne, R.~Morandotti, M.~Volatier-Ravat, V.~Aimez,
  G.~A. Siviloglou, and D.~N. Christodoulides.
\newblock Observation of $\mathcal{P}\mathcal{T}$-symmetry breaking in complex
  optical potentials.
\newblock {\em Phys. Rev. Lett.}, 103:093902, 2009.

\bibitem{Mei10}
Mei~C. Zheng, Demetrios~N. Christodoulides, Ragnar Fleischmann, and Tsampikos
  Kottos.
\newblock $\mathcal{PT}$ optical lattices and universality in beam dynamics.
\newblock {\em Phys. Rev. A}, 82:010103, 2010.

\bibitem{Hamid12}
Hamidreza Ramezani, Tsampikos Kottos, Vassilios Kovanis, and Demetrios~N.
  Christodoulides.
\newblock Exceptional-point dynamics in photonic honeycomb lattices with
  $\mathcal{PT}$ symmetry.
\newblock {\em Phys. Rev. A}, 85:013818, 2012.

\bibitem{Longhi14}
Stefano Longhi and Giuseppe Della~Valle.
\newblock Optical lattices with exceptional points in the continuum.
\newblock {\em Phys. Rev. A}, 89:052132, 2014.

\bibitem{Ge14}
Li~Ge and A.~Douglas Stone.
\newblock Parity-time symmetry breaking beyond one dimension: The role of
  degeneracy.
\newblock {\em Phys. Rev. X}, 4:031011, 2014.

\bibitem{Cerjan16}
Alexander Cerjan, Aaswath Raman, and Shanhui Fan.
\newblock Exceptional contours and band structure design in parity-time
  symmetric photonic crystals.
\newblock {\em arXiv}, (arXiv:1601.05489), 2016.

\bibitem{Berry03}
Michael~V. Berry.
\newblock Mode degeneracies and the petermann excess-noise factor for unstable
  lasers.
\newblock {\em Journal of Modern Optics}, 50(1):63--81, 2003.

\bibitem{Adi16}
A.~Pick, B.~Zhen, O.~D. Miller, C.~W. Hsu, F.~Hernandez, A.~W. Rodriguez,
  M.~Solja\v{c}i\'{c}, and S.~G. Johnson.
\newblock General theory of spontaneous emission near exceptional points.
\newblock {\em arXiv}, 1604.06478, 2016.

\bibitem{Dembowski03}
C.~Dembowski, B.~Dietz, H.-D. Gr\"af, H.~L. Harney, A.~Heine, W.~D. Heiss, and
  A.~Richter.
\newblock Observation of a chiral state in a microwave cavity.
\newblock {\em Phys. Rev. Lett.}, 90:034101, 2003.

\bibitem{Zin11}
Zin Lin, Hamidreza Ramezani, Toni Eichelkraut, Tsampikos Kottos, Hui Cao, and
  Demetrios~N. Christodoulides.
\newblock Unidirectional invisibility induced by
  $\mathcal{P}\mathcal{T}$-symmetric periodic structures.
\newblock {\em Phys. Rev. Lett.}, 106:213901, 2011.

\bibitem{Feng13}
Liang Feng, Ye-Long Xu, William~S. Fegadolli, Ming-Hui Lu, Jos{\'e}E.~B.
  Oliveira, Vilson~R. Almeida, Yan-Feng Chen, and Axel Scherer.
\newblock Experimental demonstration of a unidirectional reflectionless
  parity-time metamaterial at optical frequencies.
\newblock {\em Nat Mater}, 12(2):108--113, 2013.

\bibitem{Liertzer12}
M.~Liertzer, Li~Ge, A.~Cerjan, A.~D. Stone, H.~E. T\"ureci, and S.~Rotter.
\newblock Pump-induced exceptional points in lasers.
\newblock {\em Phys. Rev. Lett.}, 108:173901, 2012.

\bibitem{Hodaei14}
Hossein Hodaei, Mohammad-Ali Miri, Matthias Heinrich, Demetrios~N.
  Christodoulides, and Mercedeh Khajavikhan.
\newblock Parity-time{\textendash}symmetric microring lasers.
\newblock {\em Science}, 346(6212):975--978, 2014.

\bibitem{Feng14}
Liang Feng, Zi~Jing Wong, Ren-Min Ma, Yuan Wang, and Xiang Zhang.
\newblock Single-mode laser by parity-time symmetry breaking.
\newblock {\em Science}, 346(6212):972--975, 2014.

\bibitem{Gao15}
T.~Gao, E.~Estrecho, K.~Y. Bliokh, T.~C.~H. Liew, M.~D. Fraser, S.~Brodbeck,
  M.~Kamp, C.~Schneider, S.~Hofling, Y.~Yamamoto, F.~Nori, Y.~S. Kivshar, A.~G.
  Truscott, R.~G. Dall, and E.~A. Ostrovskaya.
\newblock Observation of non-hermitian degeneracies in a chaotic
  exciton-polariton billiard.
\newblock {\em Nature}, 526(7574):554--558, 2015.

\bibitem{Graefe08}
E~M Graefe, U~G{\"u}nther, H~J Korsch, and A~E Niederle.
\newblock A non-hermitian $\mathcal{P}\mathcal{T}$ symmetric bose--hubbard
  model: eigenvalue rings from unfolding higher-order exceptional points.
\newblock {\em Journal of Physics A: Mathematical and Theoretical},
  41(25):255206, 2008.

\bibitem{Ryu12}
Jung-Wan Ryu, Soo-Young Lee, and Sang~Wook Kim.
\newblock Analysis of multiple exceptional points related to three interacting
  eigenmodes in a non-hermitian hamiltonian.
\newblock {\em Phys. Rev. A}, 85:042101, 2012.

\bibitem{Wunner15}
W~D Heiss and G~Wunner.
\newblock Resonance scattering at third-order exceptional points.
\newblock {\em Journal of Physics A: Mathematical and Theoretical},
  48(34):345203, 2015.

\bibitem{PRX16}
Kun Ding, Guancong Ma, Meng Xiao, Z.~Q. Zhang, and C.~T. Chan.
\newblock Emergence, coalescence, and topological properties of multiple
  exceptional points and their experimental realization.
\newblock {\em Phys. Rev. X}, 6:021007, 2016.

\bibitem{Note1}
By accidental degeneracy, we mean that the frequency collision is neither
  induced nor protected by any underlying point symmetry, but rather is
  entirely fashioned out of the detailed morphology of the photonic unit cell,
  designed via brute-force topology optimization techniques.

\bibitem{CTChan12}
Jun Mei, Ying Wu, C.~T. Chan, and Zhao-Qing Zhang.
\newblock First-principles study of dirac and dirac-like cones in phononic and
  photonic crystals.
\newblock {\em Phys. Rev. B}, 86:035141, 2012.

\bibitem{Sakoda12}
Kazuaki Sakoda.
\newblock Proof of the universality of mode symmetries in creating photonic
  dirac cones.
\newblock {\em Opt. Express}, 20(22):25181--25194, 2012.

\bibitem{Jensen11}
J.S. Jensen and O.~Sigmund.
\newblock Topology optimization for nano-photonics.
\newblock {\em Laser and Photonics Reviews}, 5(2):308--321, 2011.

\bibitem{Piggott15}
Alexander~Y. Piggott, Jesse Lu, Konstantinos~G. Lagoudakis, Jan Petykiewicz,
  Thomas~M. Babinec, and Jelena Vuckovic.
\newblock Inverse design and demonstration of a compact and broadband on-chip
  wavelength demultiplexer.
\newblock {\em Nature Photonics}, 9:374--377, 2015.

\bibitem{Shen15}
Bing Shen, Peng Wang, and Rajesh Menon.
\newblock An integrated-nanophotonics polarization beamsplitter with 2.4 x 2.4
  um2 footprint.
\newblock {\em Nature Photonics}, 9:378--382, 2015.

\bibitem{Liang13}
Xiangdong Liang and Steven~G. Johnson.
\newblock Formulation for scalable optimization of microcavities via the
  frequency-averaged local density of states.
\newblock {\em Opt. Express}, 21(25):30812--30841, Dec 2013.

\bibitem{Zin16}
Zin Lin, Xiangdong Liang, Marko Lon\v{c}ar, Steven~G. Johnson, and Alejandro~W.
  Rodriguez.
\newblock Cavity-enhanced second-harmonic generation via nonlinear-overlap
  optimization.
\newblock {\em Optica}, 3(3):233--238, Mar 2016.

\bibitem{Svanberg02}
Krister Svanberg.
\newblock A class of globally convergent optimization methods based on
  conservative convex separable approximations.
\newblock {\em SIAM Journal on Optimization}, pages 555--573, 2002.

\bibitem{Scheel99}
S.~Scheel, L.~Kn\"oll, and D.-G. Welsch.
\newblock Spontaneous decay of an excited atom in an absorbing dielectric.
\newblock {\em Phys. Rev. A}, 60:4094--4104, 1999.

\bibitem{Petermann1979}
K.~Petermann.
\newblock Calculated spontaneous emission factor for double-heterostructure
  injection lasers with gain-induced waveguiding.
\newblock {\em IEEE J. Quant. Elect.}, 15(7):566--570, 1979.

\bibitem{berry2003mode}
M.~V. Berry.
\newblock Mode degeneracies and the petermann excess-noise factor for unstable
  lasers.
\newblock {\em J. of mod. opt.}, 50(1):63--81, 2003.

\bibitem{barnett1996sum}
S.~M. Barnett and R.~Loudon.
\newblock Sum rule for modified spontaneous emission rates.
\newblock {\em Phys. Rev. Lett.}, 77(12):2444, 1996.

\bibitem{Borisov98}
R.A. Borisov, G.N. Dorojkina, N.I. Koroteev, V.M. Kozenkov, S.A. Magnitskii,
  D.V. Malakhov, A.V. Tarasishin, and A.M. Zheltikov.
\newblock Fabrication of three-dimensional periodic microstructures by means of
  two-photon polymerization.
\newblock {\em Applied Physics B}, 67(6):765--767, 1998.

\bibitem{Lewis15}
Anders Clausen, Fengwen Wang, Jakob~S. Jensen, Ole Sigmund, and Jennifer~A.
  Lewis.
\newblock Topology optimized architectures with programmable poisson's ratio
  over large deformations.
\newblock {\em Advanced Materials}, 27(37):5523--5527, 2015.

\bibitem{Vukusic16}
Caroline Pouya, Johannes T.~B. Overvelde, Mathias Kolle, Joanna Aizenberg,
  Katia Bertoldi, James~C. Weaver, and Pete Vukusic.
\newblock Characterization of a mechanically tunable gyroid photonic crystal
  inspired by the butterfly parides sesostris.
\newblock {\em Advanced Optical Materials}, 4(1):99--105, 2016.

\bibitem{Taflove2013}
A.~Taflove, A.~Oskooi, and S.~G. Johnson.
\newblock {\em Advances in FDTD Computational Electrodynamics: Photonics and
  Nanotechnology}.
\newblock Artech House, 2013.

\bibitem{Arfken2006}
G.~B. Arfken and H.~J. Weber.
\newblock {\em Mathematical Methods for Physicists}.
\newblock Elsevier Academic Press, 2006.

\bibitem{mailybaev1999singularities}
A.~Mailybaev and A.~P. Seyranian.
\newblock On singularities of a boundary of the stability domain.
\newblock {\em SIAM Journal on Matrix Analysis and Applications},
  21(1):106--128, 1999.

\bibitem{demange2011signatures}
G.~Demange and E-M Graefe.
\newblock Signatures of three coalescing eigenfunctions.
\newblock {\em Journal of Physics A: Mathematical and Theoretical},
  45(2):025303, 2011.

\bibitem{Seyranian2003}
A.~P. Seyranian and A.~A. Mailybaev.
\newblock {\em Multiparameter Stability Theory With Mechanical Applications}.
\newblock World Scientific Publishing, 2003.

\end{thebibliography}


\begin{thebibliography}{10}

\bibitem{Jensen11}
J.S. Jensen and O.~Sigmund.
\newblock Topology optimization for nano-photonics.
\newblock {\em Laser and Photonics Reviews}, 5(2):308--321, 2011.

\bibitem{Svanberg02}
Krister Svanberg.
\newblock A class of globally convergent optimization methods based on
  conservative convex separable approximations.
\newblock {\em SIAM Journal on Optimization}, pages 555--573, 2002.

\bibitem{Liang13}
Xiangdong Liang and Steven~G. Johnson.
\newblock Formulation for scalable optimization of microcavities via the
  frequency-averaged local density of states.
\newblock {\em Opt. Express}, 21(25):30812--30841, Dec 2013.

\bibitem{CTChan12}
Jun Mei, Ying Wu, C.~T. Chan, and Zhao-Qing Zhang.
\newblock First-principles study of dirac and dirac-like cones in phononic and
  photonic crystals.
\newblock {\em Phys. Rev. B}, 86:035141, 2012.

\bibitem{Zhen15}
Bo~Zhen, Chia~Wei Hsu, Yuichi Igarashi, Ling Lu, Ido Kaminer, Adi Pick,
  Song-Liang Chua, John~D. Joannopoulos, and Marin Soljacic.
\newblock Spawning rings of exceptional points out of dirac cones.
\newblock {\em Nature}, 525(7569):354--358, 09 2015.

\bibitem{Adi16}
A.~Pick, B.~Zhen, O.~D. Miller, C.~W. Hsu, F.~Hernandez, A.~W. Rodriguez,
  M.~Solja\v{c}i\'{c}, and S.~G. Johnson.
\newblock General theory of spontaneous emission near exceptional points.
\newblock {\em arXiv}, 1604.06478, 2016.

\bibitem{Petermann1979}
K.~Petermann.
\newblock Calculated spontaneous emission factor for double-heterostructure
  injection lasers with gain-induced waveguiding.
\newblock {\em IEEE J. Quant. Elect.}, 15(7):566--570, 1979.

\bibitem{berry2003mode}
M.~V. Berry.
\newblock Mode degeneracies and the petermann excess-noise factor for unstable
  lasers.
\newblock {\em J. of mod. opt.}, 50(1):63--81, 2003.

\bibitem{Taflove2013}
A.~Taflove, A.~Oskooi, and S.~G. Johnson.
\newblock {\em Advances in FDTD Computational Electrodynamics: Photonics and
  Nanotechnology}.
\newblock Artech House, 2013.

\bibitem{Arfken2006}
G.~B. Arfken and H.~J. Weber.
\newblock {\em Mathematical Methods for Physicists}.
\newblock Elsevier Academic Press, 2006.

\bibitem{mailybaev1999singularities}
A.~Mailybaev and A.~P. Seyranian.
\newblock On singularities of a boundary of the stability domain.
\newblock {\em SIAM Journal on Matrix Analysis and Applications},
  21(1):106--128, 1999.

\bibitem{demange2011signatures}
G.~Demange and E-M Graefe.
\newblock Signatures of three coalescing eigenfunctions.
\newblock {\em Journal of Physics A: Mathematical and Theoretical},
  45(2):025303, 2011.

\bibitem{Seyranian2003}
A.~P. Seyranian and A.~A. Mailybaev.
\newblock {\em Multiparameter Stability Theory With Mechanical Applications}.
\newblock World Scientific Publishing, 2003.

\end{thebibliography}

\section{Supplementary Materials}

\title{Inverse design of a third order exceptional point via topology optimization: supplementary materials}

\author{Zin Lin$^1$}
\email{zinlin@g.harvard.edu}
\author{Adi Pick$^{2,3}$}
\author{Marko Loncar$^1$}
\author{Alejandro W. Rodriguez$^{4}$}
\affiliation{$^1$John A. Paulson School of Engineering and Applied Sciences, Harvard University, Cambridge, MA 02138}
\affiliation{$^2$Department of Physics, Harvard University, Cambridge, MA 02138}
\affiliation{$^3$Department of Mathematics, Massachusetts Institute of Technology, Cambridge, MA 02139}
\affiliation{$^4$Department of Electrical Engineering, Princeton University, Princeton, NJ, 08544}

\date{\today}

\begin{abstract}
  Write abstract here. 
\end{abstract}

\pacs{Valid PACS appear here}%
\maketitle

\end{comment}

\section{Topology optimization}

\begin{figure*}[t!]
\centering
\includegraphics[width=1.2\columnwidth]{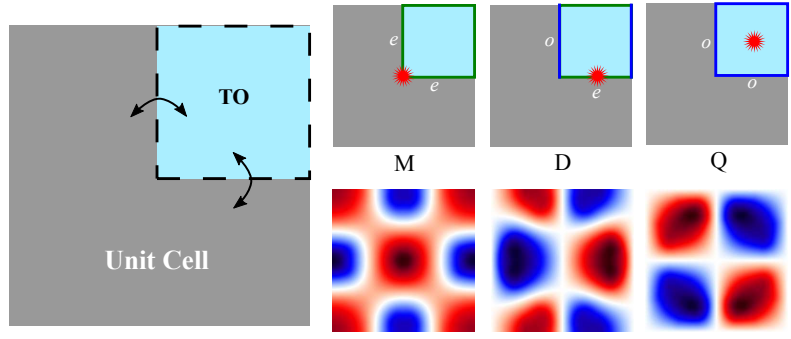}
\caption{Left: Schematic showing the design region (top-right
  quadrant) for topology optimization (TO) of a photonic crystal (PhC)
  unit cell at the $\Gamma$ point. The design region is extended by
  mirror reflections to the remaining quadrants of the unit
  cell. Right: The position of a point source (bright red dot) as well
  as choice of even ($e$) or odd ($o$) boundary conditions which
  determine the nodal structure of the resulting
  modes.~\label{fig:sfig1}}
\end{figure*}

A typical topology optimization problem in photonics goes as
follows. The objective is to maximize or minimize a given objective
function $f$ subject to certain constraints $g$ over a set of free
variables or degrees of freedom (DOF):
\begin{align}
  \text{max}/\text{min}\, &f(\epsbar_\alpha) \\
  &g(\epsbar_\alpha) \le 0 \\
  &0 \le \epsbar_\alpha \le 1
\end{align}
where the DOF are the normalized dielectric constants $\epsbar_\alpha
\in [0, 1]$ assigned to each pixel or voxel (indexed $\alpha$) in a
specified volume. Note that, in general, the index $\alpha$ denote
Cartesian components $\{\alpha_x, \alpha_y, \alpha_z\}$, such that in
a finite-difference grid, $\epsbar_\alpha = \epsbar(\vecr_\alpha) =
\epsbar(\alpha_x \Delta x, \alpha_y \Delta y, \alpha_z \Delta
z)$. Depending on the choice of background (bg) and structural
materials, $\epsbar_\alpha$ is mapped onto position-dependent
dielectric constant via $\e_\alpha = \lp \e - \e_\text{bg} \rp
\epsbar_\alpha + \e_\text{bg}$. Since we are interested in
fabricatable structures, we primarily focus on binary dielectrics by
avoiding intermediate values of $\epsbar$. The binarity of the system
can be enforced by penalizing the objective function or utilizing a
variety of filter and regularization
methods~\cite{Jensen11}. Typically, starting from a random initial
guess or completely uniform space, the technique discovers complex
structures automatically with the aid of powerful algorithms such as
the method of moving asymptotes (MMA)~\cite{Svanberg02}, which
typically require gradient information of the objective and constraint
functions, i.e., ${\partial f \over
  \partial \epsbar_\alpha},~{\partial g \over \partial \epsbar_\alpha}$. For an
electromagnetic problem, $f$ and $g$ are typically functions of the
electric $\vec{E}$ or magnetic $\vec{H}$ fields integrated over some
region, which are in turn solutions of Maxwell's equations under some
incident current or field. In what follows, we exploit direct solution
of the local Maxwell's equations (a partial differential equation),
\begin{align}
  \nabla \times {1 \over \mu}~\nabla \times \vec{E} -~
  \epsilon(\mathbf{r}) \omega^2 \vec{E} = i \omega \mathbf{J},
\label{eq:ME}
\end{align}
to obtain the steady-state $\vec{E}(\vecr;\w)$ in response to incident
currents $\vec{J}(\vecr,\w)$ at frequency $\w$.  While solution of
\eqref{ME} is straightforward and commonplace, the key to making
optimization problems tractable is to obtain a fast-converging and
computationally efficient adjoint formulation of the
problem~\cite{Jensen11}. Within the scope of TO, this requires
efficient calculations of the gradients ${\partial f \over \partial
  \epsbar_\alpha},~{\partial g \over \partial \epsbar_\alpha}$ at
every pixel $\alpha$, which we perform by exploiting the powerful
adjoint-variable method (AVM), described
in~\cite{Jensen11}. Essentially, instead of having to calculate the
${\partial f \over \partial \epsbar(\vecr)}$ for every spatial point
$\vecr$, AVM offers the gradient over the entire optimization region
at the cost of a \emph{single} (additional) solution of Maxwell's
equation, and is therefore key to the tractability of the optimization
process.

\section{LDOS formulation}

Recent work~\cite{Liang13} considered topology optimization of the
cavity Purcell factor by exploiting the concept of local density of
states (LDOS). In particular, the equivalence between the LDOS and
power radiated by a \emph{point} dipole can be exploited to reduce
Purcell-factor maximization problems to a series of small scattering
calculations. The objective function is chosen as
$\mathrm{max}_{\bar{\epsilon}}~f\lp \bar{\epsilon}(\mathbf{r});\w \rp
= - \mathrm{Re}\Big[\int d\mathbf{r}~\mathbf{J}^* \cdot \mathbf{E}
\Big]$, where $\mathbf{J}=\delta \lp \mathbf{r}
- \mathbf{r}_0 \rp \mathbf{\hat{e}}_j$. The gradient field ${\partial
  f \over \partial \bar{\epsilon} }$ is given by~\cite{Liang13}
\begin{align}
  {\partial f \over \partial \bar{\epsilon} (\mathbf{r}) } =
  \mathrm{Re} \Big[ i \w (\epsilon - \epsilon_\text{bg}) \mathbf{E}
  \cdot \mathbf{E} \Big]
\end{align}
A key realization in~\cite{Liang13} is that instead of maximizing the
LDOS at a single discrete frequency $\w$, a better-posed problem is
that of maximizing the frequency-averaged $f$ in the vicinity of $\w$,
denoted by $\langle f \rangle = \int d\w'~{\cal W}(\w';\w,\Gamma)
f(\w')$, where ${\cal W}$ is some weight function defined over a
specified bandwidth $\Gamma$. Using contour integration techniques,
the frequency integral can be conveniently replaced by a single
evaluation of $f$ at a complex frequency $\w + i
\Gamma$~\cite{Liang13}. For a fixed $\Gamma$, the frequency average
effectively forces the algorithm to favor minimizing $V$ over
maximizing $Q$; the latter can be enhanced over the course of the
optimization by gradually winding down $\Gamma$~\cite{Liang13}. A
major merit of this formulation is that it features a mathematically
well-posed objective as opposed to a direct maximization of the cavity
Purcell factor ${ Q \over V}$, allowing rapid convergence into
extremal solutions. Here, we note that the LDOS formulation offers a
natural elegant tool for the inverse design of any kind of resonant
mode, not just the localized cavity modes considered
in~\cite{Liang13}. In particular, it can be successfully applied for
the inverse design of extended Bloch modes in a periodic medium for an
arbitrary choice of Bloch wave vector $\mathbf{k}$. Although this work
has focused on Bloch modes at the $\Gamma$ point, we have found that
the algorithm can be employed with similar ease to design photonic
spectra at $\mathbf{k} \neq 0$.

A simple extension of the optimization formula from a single-mode
problem to the inverse design of a multi-mode degeneracy is to
maximize the minimum of a collection of LDOSs corresponding to
different $\mathbf{J}$'s at the exact same frequency $\w$. Here, the
objective assumes the form of a so-called \emph{maximin} problem:
$\mathrm{max}_{\bar{\epsilon}(\mathbf{r})} \mathrm{min}_n \Big\{ f
(\w;\mathbf{J}_n)\Big\}$, which requires solving \emph{separate}
scattering problems for the distinct sources $\mathbf{J}_n$ at the
same frequency $\w$. In practice, we replace the maximin objective
with an equivalent formulation~\cite{Jensen11}:
$\mathrm{max}~t,~\text{subject to } t-f_n \le 0$.

\section{Design of an accidental third-order Dirac degeneracy at the $\Gamma$ point}

To design a third order Dirac degeneracy (D3), we maximize the minimum
of $\Big\{ f_\text{M},~ f_\text{D},~f_\text{Q} \Big\}$ at the $\Gamma$
point, where M, D and Q denote monopolar, dipolar and quadrupolar
transverse magnetic (TM) modes ($\hat{z} \times \mathbf{E} = 0$). For
easier computations, we impose $C_{2v}$ symmetry with mirror planes at
the center of the unit cell. Note that the mirror planes are also
essential for classifying modes by their even or odd transformation
properties. In group theoretic language, the eigenmodes of the PhC at
$\Gamma$ point transform according to distinct irreducible
representations; specifically, M, D and Q modes belong to three
\emph{distinct} irreducible representations A$_1$, A$_2$ and B$_1$ of
the group $C_{2v}$. In effect, the degrees of freedom (DOF) are
restricted to one quadrant of the unit cell whereas the unique nodal
structures of M, D and Q are enforced by a careful choice of boundary
conditions as well as a judicious positioning of the point sources
$\mathbf{J}$, as shown in~\figref{sfig1}. Under these settings, the
optimization converges approximately within 500 iterations, taking
less than two hours. During optimization, we also impose filter and
penalization constraints~\cite{Jensen11} in order to avoid
intermediate $\epsilon$ values.

\begin{figure*}[ht!]
\centering
\centering \includegraphics[width=1.9\columnwidth]{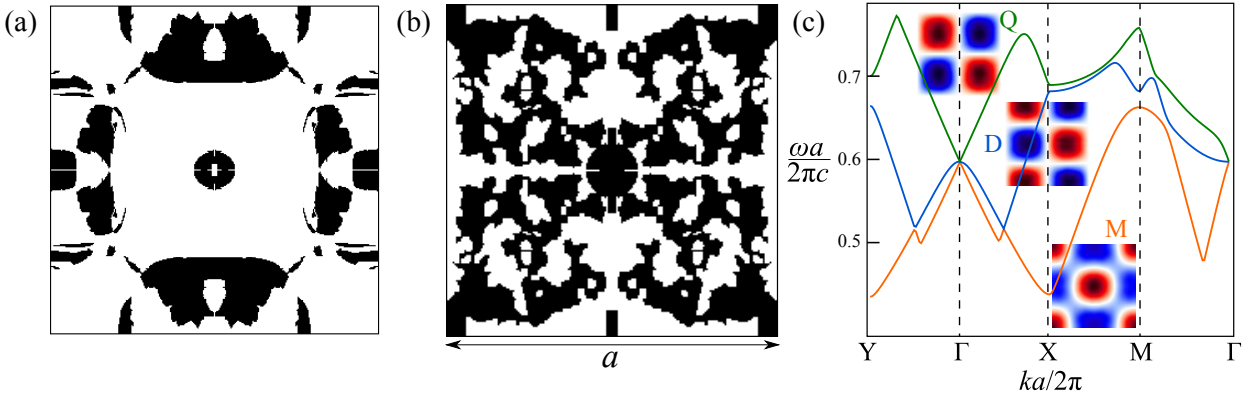}
\caption{Detailed images of inverse-designed PhC unit cells comprising
  either (a) low-index ($n=2$) or (b) high-index ($n=3.07$) dielectric
  materials in air, and supporting third-order Dirac
  degeneracies. Black/white represent dielectric/air regions. The
  lattice constants of the cells are $a=1.05~\lambda$ and
  $a=0.6~\lambda$, respectively, where $\lambda$ denotes the design
  wavelength. Each pixel has dimensions $\approx 0.003 \times 0.003
  a^2$. (c) Band structure of the high-index structure, demonstrating
  the conical, Dirac dispersion around a tri-modal Dirac degeneracy,
  involving monopolar (M), dipolar (D), and quadrupolar (Q) modes, at
  the $\Gamma$ point.~\label{fig:sfig2}}
\end{figure*}

\Figref{sfig2} shows two binary structures obtained by application of
the aforementioned optimization technique and exhibiting the desired
three-mode degeneracy to within $0.01\%$ of the designated
frequencies, $\omega_0= 2 \pi c / a$, where $a$ is the lattice
constant. While the details of the low-index structure
[\figref{sfig2}(a)] are described in the main text, here we focus on
the high-index design [\figref{sfig2}(b)], whose refractive index
$n=3.07$ and period $a=0.6~\lambda$. Noticeably, the high-index design
possesses highly connected features (few isolated components) and
should, therefore, be more readily fabricatable by conventional
methods. For instance, the dielectric constant of alumina ceramics is
$\approx 9.4$ at $10~\mathrm{GHz}$, paving the way for fabrication and
characterization of such a structure based on standard high-precision
computerized machining of suitable alumina samples at microwave
frequencies. The band structure of the high-index design exhibits a D3
of M, D and Q modes~\figref{sfig2}(b). Assuming loss
$\mathrm{Im}[\epsilon] = 0.02$ uniformly distributed throughout the
dielectric material, leading to decay rates
$\{\gamma_\text{M},\gamma_\text{D},\gamma_\text{Q}\}/\omega_0 \approx
\{9.83,9.73,9.78\} \times 10^{-4}$, we find that the EP3 occurs at
$\mathbf{k}_\text{EP3}=\{4.5,2.9\} \times 10^{-6} (2 \pi / a)$ and
results in a Petermann factor $\approx 10^8$.

\section{Design of a third-order exceptional point}

In order to better understand the dispersion properties of the
TO-designed PhC as well as to determine the existence of an EP3, we
can approximate the band structure near the $\Gamma$ point in terms of
the degenerate modes at $\Gamma$, leading to an eigenproblem
$\mathcal{H} \psi = \omega^2 \psi$ based on the $3 \times 3$
Hamiltonian~\cite{CTChan12}:
\begin{align}
\mathcal{H} = \begin{pmatrix}
\omega_0^2 & p_\text{MD} k_x & 0 \\
p^*_\text{MD} k_x & \omega_0^2 & p_\text{QD} k_y \\
0 & p^*_\text{QD} k_y & \omega_0^2
\end{pmatrix} \label{eq:eqH}
\end{align}
Note that under the approximation $\omega_0^2 - \omega^2 \approx 2
\omega_0 (\omega_0 - \omega)$ and substitution $p_{ij} = 2 \omega_0
v_{ij}$, one is led to the simplified Hamiltonian (considered in the
main text):
\begin{align}
\mathcal{H}=
\begin{pmatrix}
\omega_0 & v_\text{MD} k_x & 0 \\
v^*_\text{MD} k_x & \omega_0 & v_\text{QD} k_y \\
0 & v^*_\text{QD} k_y & \omega_0
\end{pmatrix} \label{eq:eqH2}
\end{align}
Although \eqref{eqH2} is easier to work with for deriving closed-form
analytical expressions, to achieve better accuracy our predictions in
the main text and discussion below are based on \eqref{eqH}.

\begin{figure*}[ht!]
\centering
\includegraphics[width=1.7\columnwidth]{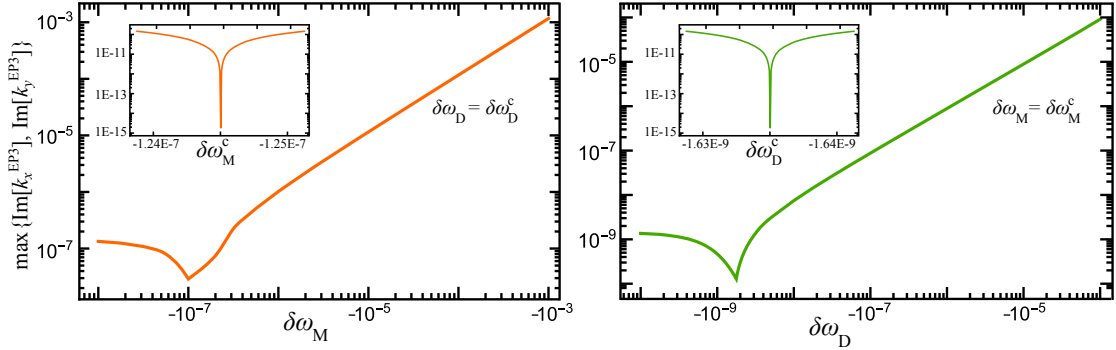}
\caption{Left: Imaginary part of $\mathbf{k}^\text{EP3}$ as a function
  of gap parameter $\delta \omega_\text{M}$, with $\delta
  \omega_\text{D}$ fixed at the critical value $\delta
  \omega_\text{D}^\text{c}$. Note the approximately linear scaling of
  $\mathrm{Im}[\mathbf{k}^\text{EP3}] \sim \delta \omega_\text{M}$
  away from the critical point. The inset magnifies the vicinity of
  $\delta \omega_\text{M}^\text{c}$, the critical gap value at which
  $\mathrm{Im}[\mathbf{k}^\text{EP3}]$ vanishes. Right: Same as left
  but exploring variations of $\delta \omega_\text{D}$ when $\delta
  \omega_\text{M}$ is fixed at $\delta
  \omega_\text{M}^\text{c}$.~\label{fig:sfig3}}
\end{figure*}

The introduction of a small $\mathrm{Im}[\epsilon]$ yields the
following non-Hermitian Hamiltonian:
\begin{align}
\mathcal{H}' = \begin{pmatrix}
(\omega_0 + i \gamma_\text{M})^2 & p'_\text{MD} k_x & 0 \\
-p'_\text{MD} k_x & (\omega_0 + i \gamma_\text{D})^2 & p'_\text{QD} k_y \\
0 & -p'_\text{QD} k_y & (\omega_0 + i \gamma_\text{Q})^2
\end{pmatrix} \label{eq:eqnH}
\end{align}
Note that for sufficiently small $\mathrm{Im}[\epsilon]$, $p'_{ij}
\approx p_{ij}$ and $-p'_{ij} \approx p^*_{ij}$ and that the form
of~\eqref{eqnH} maintains reciprocity since
$\mathcal{H}^\text{T}(\mathbf{k}) = \mathcal{H}(-\mathbf{k})$. For
simplicity of notation, we will drop the prime with the understanding
that any reference to $\mathcal{H}$ from here on refers
to~\eqref{eqnH}. The mode-mixing parameters $p_{ij}$ can be computed
from overlap integrals between the degenerate modes at the $\Gamma$
point~\cite{CTChan12}. In particular, in the case of the low-index
design ($n=2$, see main text) where we have chosen
$\mathrm{Im}[\epsilon] = 0.005$, we obtain $\gamma/\omega_0 \sim
10^{-3}$, $p_\text{MD} \approx 5.9i/(2 \pi)$ and $p_\text{QD} \approx
5.5i/(2 \pi)$. With these parameters in hand, we can determine the
location of the EP3 by numerically solving \eqref{det1}--\eqref{det2}
in the main text, resulting in the aforementioned values of
$k_x^\text{EP3} \approx 7\times 10^{-5}~({2 \pi \over \lambda})$ and
$k_y^\text{EP3} \approx 1.8\times 10^{-5}~({2 \pi \over \lambda})$.

While the topology-optimized binary design exhibits a tri-modal
degeneracy to an accuracy of $\lesssim 0.01\%$, we find that in order
to access the EP3, further fine-tuning is necessary as is generally
the case for parameter-sensitive exceptional points~\cite{Zhen15,
  Adi16}. In particular, for a fixed $\omega_\text{Q}=\omega_0$, small
deviations from some critical frequencies $\omega_\text{M}^\text{c}$
and $\omega_\text{D}^\text{c}$ introduces a small imaginary part to
$\mathbf{k}_\text{EP3}$. \Figref{sfig3} quantifies the magnitude of
$\mathrm{Im}[\mathbf{k}_\text{EP3}]$ as a function of two bandgap
parameters $\delta \omega_\text{M}$ and $\delta \omega_\text{D}$,
defined such that $\omega_\text{M} = \omega_0 + \delta
\omega_\text{M}$ and $\omega_\text{D} = \omega_0 + \delta
\omega_\text{D}$. As observed, the imaginary part of
$\mathbf{k}_\text{EP3}$ vanishes when $\delta \omega_\text{M} = \delta
\omega_\text{M}^\text{c} \sim -10^{-7}$ and $\delta \omega_\text{D} =
\delta \omega_\text{D}^\text{c} \sim -10^{-9}$, signaling the
appearance of a EP3 on the real $\mathbf{k}$ plane. While there are
many post-fabrication fine-tuning techniques (such as oxidation,
thermal, free-carrier, or laser tuning), in our numerical experiment,
we simply fine-tune a few strategic pixels in the PhC design to vary
$\omega_\text{M}$ and $\omega_\text{D}$ while keeping
$\omega_\text{Q}$ fixed, repeatedly solving the full Maxwell
eigenproblems until the Petermann Factor $\sim 10^9$.

\section{Green's Function at a third-order EP}

Non-orthogonality of the modes in open resonators can lead to
significantly enhanced spontaneous emission
rates~\cite{Petermann1979}. This effect becomes most pronounced near
exceptional points~\cite{berry2003mode}, where the modes become
self-orthogonal.  The figure of merit for computing spontaneous
emission rates is the local density of states (LDOS), which is
proportional to the imaginary part of the Green's function
(GF)~\cite{Taflove2013}.  Near non-degenerate resonances, the GF can
be expressed using the standard modal expansion
formula~\cite{Arfken2006}:
\begin{align}
  G = \sum_i\frac{1}{\omega^2-\omega^2_i}\cdot
  \frac{\Psi^\text{R}_i (\Psi^\text{L}_i)^T}{(\Psi^\text{L}_i)^T \Psi^\text{R}_i}.
\label{eq:non-degenerate}
\end{align}
The right eigenvectors $\Psi^\text{R}_i$ and eigenvalues $\omega_i$
are outgoing solutions of Maxwell's equations or, more explicitly,
satisfy the eigenvalue problem: $A\Psi^\text{R}_i =
\omega^2_i\Psi^\text{R}_i$. Here, $A$ is Maxwell's operator
$\varepsilon^{-1}\nabla\times\nabla\times$ and $\varepsilon$ is the
dielectric permittivity.  Left eigemodes are eigenvectors of the
transposed operator $A^T\Psi_i^\text{L} = \omega^2_i\Psi_i^\text{L}$,
where $A^T \equiv \nabla\times\nabla\times\varepsilon^{-1}$.  The
derivation of \eqref{non-degenerate} relies on the assumption that the
set of eigenvectors of $A$ spans the Hilbert space, which breaks down
at EPs due to the coalescence of both the eigenvalues and
eigenvectors. In what follows, we derive an eigenvalue expansion
formula for the GF that is valid at third-order exceptional points
(EP3). Our derivation follows three main steps (as in~\cite{Adi16}):
First, we use perturbation theory to express the eigenvalues
$\omega_i$ and eigenmodes $\Psi_i^\text{L}$ near the EP in terms of
the degenerate eigenvalue and Jordan-chain vectors and an associated
perturbative parameter. We then substitute these expressions into
\eqref{non-degenerate}. Lastly, we take the limit as one approaches
the EP.

Let the Maxwell operator $A(p)$ be a parameter-dependent operator
supporting a EP3 at $p=0$. The Jordan chain vectors of
$A_\text{EP3}\equiv A(0)$ satisfy the
relations~\cite{mailybaev1999singularities,demange2011signatures}:
\begin{align}
  & A_\text{EP3}\Psi^\text{R}_\text{EP3} = \omega^2_\text{EP3}\Psi^\text{R}_\text{EP3} \\
  & A_\text{EP3}\Phi^\text{R}_\text{I} = \omega^2_\text{EP3}\Phi^\text{R}_\text{I} + \Psi^\text{R}_\text{EP3} \\
  & A_\text{EP3}\Phi^\text{R}_\text{II} =
  \omega^2_\text{EP3}\Phi^\text{R}_\text{II} + \Phi^\text{R}_\text{I},
\end{align}
with the duals obtained by letting $A \to A^T$ and $R \to L$, leading
to the following orthogonality relations:
\begin{align}
&(\Psi^\text{L}_\text{EP3})^T \Psi^\text{R}_\text{EP3} = 0 \nonumber\\
&(\Phi^\text{L}_\text{I})^T\Psi^\text{R}_\text{EP3} =(\Psi^\text{L}_\text{EP3})^T \Phi^\text{R}_\text{I} = 0 \nonumber \\
&(\Psi^\text{L}_\text{EP3})^T) \Phi^\text{R}_\text{II} =(\Phi^\text{L}_\text{II})^T\Psi^\text{R}_\text{EP3} = (\Phi^\text{L}_\text{I})^T\Phi^\text{R}_\text{I}
\end{align}
In order to uniquely define the above chain vectors, we choose the
additional normalization conditions:
\begin{align}
&(\Psi^\text{L}_\text{EP3})^T\Phi^\text{R}_\text{II} =(\Phi^\text{L}_\text{II})^T\Psi^\text{R}_\text{EP3} = (\Phi^\text{L}_\text{I})^T\Phi^\text{R}_\text{I}= 1 \nonumber \\
&(\Phi^\text{L}_\text{II})^T\Phi^\text{R}_\text{I}  =  (\Phi^\text{L}_\text{I})^T\Phi^\text{R}_\text{II}0 = 0 \nonumber \\
&(\Phi^\text{L}_\text{II})^T\Phi^\text{R}_\text{II} =0.
\end{align}

When the LDOS is dominated by three non-degenerate resonances, one can
approximate the full GF via \eqref{non-degenerate} by keeping only
three terms in the sum.  (This requires that the three resonances
$\omega_i$ be spectrally separated from the rest of the eigenvalues
and that $G$ be evaluated at $\omega \approx \mathrm{Re}[\omega_i]$).
Near the EP, $A(p)$ can can be written as~\cite{Adi16}:
\begin{align}
A(p) = A_\text{EP3} + pA_1 + p^2A_2 + \hdots,
\label{eq:expandA}
\end{align}
from which it follows that the eigenvalues and eigenvectors of $A(p)$
can be expanded in Puiseux series~\cite{Seyranian2003},
\begin{align}
&\omega^2_i = \omega^2_\text{EP3} + p^\frac{1}{3}\omega^2_1 + p^\frac{2}{3}\omega^2_2 +  p\, \omega^2_3 + p^\frac{4}{3}\omega^2_4\hdots
\label{eq:expandLambda}\\
&\Psi_i = \Psi^\text{R}_\text{EP3} + p^\frac{1}{3}\Psi_1 + p^\frac{2}{3}\Psi_2 + p\, \Psi_3 + p^\frac{4}{3}\Psi_4
\label{eq:expandU}\hdots
\end{align} 
reducing to the eigenvalues and eigenvectors of $A_\text{EP3}$ in the
limit $p\rightarrow0$.  (Note that one can write similar expressions
for the left eigenvectors.)  Using \eqssref{expandA}{expandU} and
taking the limit as $p\rightarrow0$, we arrive at \eqref{G_ep3} in the
main text, describing the GF at a EP3. Note that in order to obtain
the correct limit, one needs to keep terms up to
$\mathcal{O}(p^{\frac{5}{3}})$ in \eqssref{expandA}{expandU}.

%\bibliographystyle{unsrt}
%\bibliography{ep3,bibliographyEP3}

%\end{document}

\end{document}